\documentclass[letterpaper,twocolumn,10pt]{article}
\usepackage{usenix,epsfig,endnotes}

\usepackage{bbding}
\usepackage{wasysym} 
\usepackage{multirow}
\usepackage{clipboard} 
\usepackage{pdflscape} 
\usepackage{csquotes}
\usepackage{changes}
\usepackage{color,soul}
\usepackage{hyperref}

\usepackage{changes}
\definechangesauthor[name={Sirko},color=green]{s}

\usepackage{adjustbox}
\usepackage{array}
\newcolumntype{R}[2]{%
    >{\adjustbox{angle=#1,lap=\width-(#2)}\bgroup}%
    c%
    <{\egroup}%
}
\newcommand*\rot{\multicolumn{1}{R{60}{1em}}}

\usepackage[inline]{enumitem}

\usepackage{scalerel}
\usepackage{tikz}
\usetikzlibrary{svg.path}
\definecolor{orcidlogocol}{HTML}{A6CE39}
\tikzset{
  orcidlogo/.pic={
    \fill[orcidlogocol] svg{M256,128c0,70.7-57.3,128-128,128C57.3,256,0,198.7,0,128C0,57.3,57.3,0,128,0C198.7,0,256,57.3,256,128z};
    \fill[white] svg{M86.3,186.2H70.9V79.1h15.4v48.4V186.2z}
    svg{M108.9,79.1h41.6c39.6,0,57,28.3,57,53.6c0,27.5-21.5,53.6-56.8,53.6h-41.8V79.1z M124.3,172.4h24.5c34.9,0,42.9-26.5,42.9-39.7c0-21.5-13.7-39.7-43.7-39.7h-23.7V172.4z}
    svg{M88.7,56.8c0,5.5-4.5,10.1-10.1,10.1c-5.6,0-10.1-4.6-10.1-10.1c0-5.6,4.5-10.1,10.1-10.1C84.2,46.7,88.7,51.3,88.7,56.8z};
  }
}
\newcommand\orcidicon[1]{\href{https://orcid.org/#1}{\mbox{\scalerel*{
  \begin{tikzpicture}[yscale=-1,transform shape]
  \pic{orcidlogo};
  \end{tikzpicture}
}{|}}}}

\begin{document}

\date{}

\title{\Large \bf Astronomical Pipeline Provenance: A Use Case Evaluation }

\author{
{\rm Michael A. C. Johnson$^1$ $^2$ \Envelope
\orcidicon{0000-0002-5566-6147}}
\and
{\rm Marcus Paradies$^1$ \orcidicon{0000-0002-5743-6580}}
\and
{\rm Marta Dembska$^1$
\orcidicon{0000-0002-8180-1525}}
\and
{\rm Kristen Lackeos$^2$ \orcidicon{0000-0002-6554-3722}} 
\and
{\rm Hans-Rainer Kl\"ockner$^2$ \orcidicon{0000-0002-0648-2704}}
\and 
{\rm David J. Champion$^2$ \orcidicon{0000-0003-1361-7723}}
\and
{\rm Sirko Schindler$^1$ \orcidicon{0000-0002-0964-4457}}
\and 
$^1$Institute of Data Science (DLR)\quad 
$^2$ Max Planck Institute for Radio Astronomy \\ 
\Envelope mjohnson@mpifr-bonn.mpg.de
}

\maketitle

\subsection*{Abstract}

In this decade astronomy is undergoing a paradigm shift to handle data from next generation observatories such as the Square Kilometre Array (SKA) or the Vera C. Rubin Observatory (LSST). Producing real time data streams of up to 10 TB/s and data products of the order of 600 Pbytes/year, the SKA will be the biggest civil data producing machine of the world that demands novel solutions on how these data volumes can be stored and analysed. Through the use of complex, automated pipelines the provenance of this real time data processing is key to establish confidence within the system, its final data products, and ultimately its scientific results. 

The intention of this paper is to lay the foundation for making an automated provenance generation tool for astronomical/data-processing pipelines. We therefore present a use case analysis, specific to the astronomical needs which addresses the issues of trust and reproducibility as well as other ulterior use cases which are of interest to astronomers. This analysis is subsequently used as the basis to discuss the requirements, challenges, and opportunities involved in designing both the tool and the associated provenance model.



\section{Introduction}

Technological advances in telescopes, detectors and computer analysis have improved the data collecting capabilities of astronomers dramatically.
This enabled a paradigm shift to change the main method of data collection from targeting specific objects to large scale surveys, which observe large regions of the sky.
The next generation of survey telescopes such as the Vera C. Rubin Observatory (LSST) or the Square Kilometre Array (SKA) will collect data in the petabyte and exabyte scale regimes, respectively. These surveys will undoubtedly provide a rich resource for data mining astronomers; however, their scale also presents challenges for data management, data analysis, and provenance generation. These challenges stem from not only the size of the datasets themselves but also the necessary complexity and distributed nature of the required pipelines.

The astronomical pipelines which are designed to analyse these datasets are typically created within scripting languages such as Python.
Some notable examples of this astronomical, pythonic data processing include: the LSST stack \cite{juric2015lsst} which is responsible for processing data from the Vera C. Rubin Observatory; and the Common Astronomy Software Application (CASA) \cite{jaeger2008common}. 

Provenance recording within any scientific discipline is instrumental in establishing trust and reproducibility in the data processing and data products. 
Despite the benefits and the current availability of provenance generation systems designed for scientific pipelines (e.g., Chimera \cite{foster2002chimera}, Taverna \cite{oinn2004taverna}, and Kepler \cite{altintas2006provenance}) recording and distribution of provenance within astronomical pipelines is not standard practice.
Implementation of these systems require users to build their pipelines within the respective workflow management systems. 
Astronomers have a reluctance to meet this requirement due to their proficiency within their native scripting languages.
In order to encourage adoption of any tool for this community, it is therefore important to meet them where they are and to not require domain specific provenance knowledge.

Examples of provenance generation tools specifically for scripting languages include NoWorkflow \cite{murta2014noworkflow} (Python only) and YesWorkflow \cite{mcphillips2015yesworkflow} (language independent).
NoWorklow automatically generates provenance directly from Python scripts, however it is unable to scale to meet the demands of astronomical datasets. 
YesWorkflow is a language independent implementation which gathers provenance according to manual annotations, putting the responsibility on its users. 


The increasing importance of provenance within astronomy, the current timeliness of the approaching next generation of telescopes, and the current lack of a suitable provenance generation tool for the astronomical community are the three main motivations behind the VAMPIRA\footnote{Acronym of the German description: {\tt V}erwaltung, {\tt A}uswertung und {\tt M}odellierung von {\tt P}rovenienz {\tt I}nformationen aus {\tt R}echenintensiven wissenschaftlichen {\tt A}rbeitsabl\"aufen.} project which aims to facilitate the automated generation of provenance for astronomical pipelines. This paper outlines the first steps towards the VAMPIRA project and details an analysis of use cases which are of interest to astronomers and can be solved using provenance. This use case analysis was subsequently used to determine the set of possible requirements for a provenance model tailored to astronomical datasets. Furthermore, we outline the challenges and opportunities of provenance generation within astronomy.

The structure of the paper is as follows: \autoref{sec:usecases} details the use case analysis and their requirements; \autoref{sec:OandC} describes the opportunities and challenges of provenance recording for astronomical datasets; Section \autoref{sec:RW} discusses the related work; \autoref{sec:conc} presents our conclusions and outlines the future work.

\section{Use Case Analysis}\label{sec:usecases}

The starting point for use case (UC) generation was through discussion with users of two separate astronomical pipelines which were each devoted to analysing data recorded at different regions within the electromagnetic spectrum.
The first was the European Pulsar Timing Array (EPTA) timing data reduction pipeline at the Max Planck Institute for Radio astronomy \cite{lazarus2020coastguard}.
The EPTA is comprised of a series of radio telescopes and one of the main objects is the study of gravitational waves. 
The second application was a transient detection pipeline which analysed the optical full frame images released by the Kepler Space Telescope.
The main objective of this application was the population study of variable Galactic objects. 
In each application, the pipeline is entirely written within Python. 
The UC discussion was further developed beyond these pipelines through discussion with several domain experts.

\subsection{Methodology}

The following methodology was applied to construct the UCs described within this paper.

\begin{itemize}[itemsep=0pt]

\item Users from each application were asked to identify provenance-related UCs which were relevant within their domain. 

\item The users were further asked to identify problems they encounter within pipeline creation and or use, as well as any pipeline information which they would like to determine. Where possible, these contributions were also constructed into UCs.

\item UCs, from both aforementioned sources, which were similar in function from a provenance standpoint were collated and a generalised description was written for them.

\item The users verified that the description accurately encompassed their UCs and an example scenario was added to each general description.

\item The requirements that each UC imposed on the provenance system were determined and the UC analysis was further refined to only include UCs which contained distinct sets of requirements. 

\item This refined set of UCs was then grouped depending on the intention of their function.

\end{itemize} 

Through this analysis, it was evident that the uses for provenance from astronomical pipelines are not solely limited to trust and repeatability.
Fine-grained provenance information can also enable ulterior UCs, such as, prediction of pipeline components or parameters, cost vs. gain analysis, or anomaly detection.

\subsection{Terminology}

Throughout this discussion, we make extensive use of the following terms.
In order to prevent misunderstanding, we provide their definitions.

\setlist[description]{font=\normalfont\itshape\space}
\begin{description}[itemsep=0pt,topsep=0pt]

  \item[Pipeline] - An organised collection of modules designed to analyse data.


  \item[Pipeline Run] - An individual instance of a pipeline execution

  \item[Pipeline Version] - A particular form of a pipeline containing a distinct set of modules and versions thereof

  \item[Pipeline Component] - An individual module used throughout a pipeline to generate, manipulate, or otherwise process data (e.g., a library, service or function) including versioning information

  \item[Parameter] - A value to be consumed by a component within a pipeline which alters its function 


\end{description}




\subsection*{Group A (\Copy{ucg:re-exec}{Trust and Reproducibility})}

Scientists depend on high-quality and reliable data.
Vouching for a given data product is one of the main motivations for recording provenance data.
The UCs in this group focus on how provenance can be used to promote trust in data processing and data products.

\paragraph*{UC 1 (\Copy{uc:re-exec}{Re-execution of existing pipeline})}

\noindent
The ability to re-execute a specific pipeline run is a fundamental requirement for scientific pipelines to facilitate repeatability of results or to cope with software and hardware failures during the execution.
The specific pipeline run can be described by the involved  components, data flows between them, used data sources, specific parameter configuration, and the used runtime environment(s).

\underline{\textit{Scenario:}}
An astronomer is using a pipeline to detect fast radio bursts (FRBs) within the observations of a radio telescope survey.
The pipeline detects a new FRB within the observations of the telescope. However, it is noticed that the survey included observations of same field a week prior which should also have triggered a detection. 
The astronomer re-executes the pipeline on these prior data to determine whether the non-detection was due to a technical failure during the original execution.


In order to re-execute a specific pipeline run, all its data sources and software components have to be available.
Furthermore, the data flow among these components needs to be captured along with any other parameters used.
In some cases, the execution environment also has to be re-created to mitigate interfering factors such as issues derived from switching the operating system. 
We note that data source and software versioning is an important aspect of reproducible science, but is not in the focus of this work.

\paragraph*{UC 2 (\Copy{uc:errors}{Impact of Errors})}

Most elements a pipeline uses undergo a continuous development that is independent from any such pipeline.
From time to time this evolution discovers issues that invalidate past results based on these elements.
This can include software bugs, errors within the input data, or the realisation that certain assumptions were plain out wrong.
Once such deficiencies become apparent, their impact needs to assessed and affected data products to be determined.
Provenance records can help to identify all pipelines that made use of these now known to be defunct elements and thus produced compromised data products.

\underline{\textit{Scenario I:}}
An astronomer is preparing a new catalogue of transient sources detected with a network of telescopes with increased sensitivity over previous surveys.
Each source detection triggers the saving of the associated raw data.
The astronomer discovers that a new class of ultra-fast transients is contained within the saved data.
This discovery needs to be confirmed through additional observations.
The astronomer determines the specific telescope and pipeline configurations that produced the detection in question to aid in the localisation of the detection to enable targeted follow-up observations.

\underline{\textit{Scenario II:}}
An astronomer maintains a pipeline which analyses the images from a survey telescope in real time in order to detect transient and variable objects.
The pipeline implements  transient detection software which was periodically updated.
A specific version of this software was later found to be faulty and results which used this version need to be recomputed.
The astronomer determines all data products which were processed using the faulty software version. 

\underline{\textit{Scenario III:}}
An astronomer measured the magnitude of a star within an astronomical image.
To correct for imperfections that contaminate the image, the astronomer used several standard stars (stars which have a known and constant brightness) within the image to generate a correction factor.
A star contained within the image that was previously thought to be standard has now been found to be variable.
The astronomer determines whether this star was used in the calibration process to decide whether the pipeline should be recomputed. 

The prime requirement of this UC is the unambiguous identification of any component that might cause issues.
This includes in particular all data sources, components, and parameters.
The internal data flow is then used to determine which data products are affected by the particular issue.
Furthermore, all data products have to be identifiable in order to flag them as possibly defunct.
A sophisticated system might even rely on the metadata of data sources to identify the ones affected and thus allow for more detailed queries to be posed.

\subsection*{Group B (\Copy{ucg:precution}{Prediction})}

Over time, users acquire a certain degree of experience which enables them to make educated guesses on the function of a new or changed pipeline.
However, this information is usually only gathered by running a pipeline and evaluating the results.
In face of ever increasing volumes of data and a growing number of pipelines, this approach is not sustainable.
Provenance records offer an alternative way of capturing this experience that allows for prediction of the performance of a given pipeline.

\paragraph*{UC 3 (\Copy{uc:compPref}{Prediction of Pipeline Performance})}

\noindent
Before running a new or changed pipeline, the characteristics of the results typically remain unclear.
The engineering process of pipelines is often driven by the experience and intuition of their authors rather than verifiable facts.
However, these facts can be provided by provenance records from past pipeline runs.
Augmented with metrics to describe the quality of the results, these records can be used to predict the outcome of a yet untested pipeline.
Possible metrics can be distinguished into two categories:
Domain-dependent metrics describe the quality of data products from a scientific point of view including, e.g., the completeness or accuracy of the result.
On the other hand, domain-independent metrics measure the data processing itself like execution times or processing costs.
With such metrics available for a large pool of documented pipeline runs, a system can approximate the characteristics of pipelines or components thereof without having to actually execute them.

\underline{\textit{Scenario I:}}
An astronomer is building a source detection pipeline to find the optical counterpart to a gravitational wave detection.
As the significance of finding this counterpart is very high and the rate of gravitational wave detections is low, the astronomer wants to design a pipeline which prioritises completeness over accuracy.
The astronomer analyses the relationship between completeness and pipeline composition from past source detection pipelines to determine which components and configuration is likely to yield the highest completeness.

\underline{\textit{Scenario II:}}
The astronomer has completed the source detection pipeline described above and needs to generate an estimate of the expected run time and resources required in order to request resources from their institution's compute cluster.
The astronomer queries the same set of past provenance for information on the resources that this pipeline is likely to consume. 

At the very minimum, this UC needs information about the involved components in the pipeline and a measure to predict.
At this point we do not distinguish between predicting the runtime characteristics, i.e. execution time or resources required, from predicting the quality of the resulting data products.
From a very abstract point of view, both can be seen as a target value or function that needs to be predicted.
Depending on the intention, the target measure is either given by quality metrics of the resulting data products or the resource consumption of the pipeline or parts thereof.
Several other aspects can be used to improve a prediction:
Other versions of the same pipeline alongside their runs can be used to perform case based reasoning.
More sophisticated solutions might also consider all kinds of factors that may influence the outcome, such as the data flow among components and the parameters used to call them.
Similarly, certain characteristics of the data sources like file size or number of observed sources can be used to increase the precision of the prediction.
Finally, also the runtime environment might influence performance and as such might become a valuable piece of information,

\paragraph*{UC 4 (\Copy{uc:storage}{Storage vs. Recomputation Analysis})}

Different pipelines may partially overlap in their components like sharing an initial cleaning step before proceeding to the actual computations.
For reoccurring portions of pipelines, it can be worthwhile to store intermediate data products and reuse them wherever necessary.
However, the decision should not rely on the number of reuses alone, but include other factors like cost of storage or resources required for recomputation.
Provenance data can provide the necessary information to make an informed decision on which intermediate data products are worth to keep beyond a particular pipeline run.

\underline{\textit{Scenario:}}
An astronomer builds a pipeline to analyse the data products produced by a survey telescope.
The aim of the pipeline is the detection of astronomical objects, however it also contains a number of cleaning and pre-processing steps which are performed prior to detection. 
Each of these steps produce intermediate data products such as cleaned images which are all stored by default for use by future pipelines.
As time progresses, more data is collected and the astronomer develops further pipelines to analyse the period and magnitude of the detected objects which utilise the intermediate data products, such as the cleaned images.
The number of intermediate data products increases to fill the available storage resources so the astronomer must decide which to compute on-the-fly and which to cache for later reuse.
The astronomer determines the frequency of use of each intermediate data products as well as their associated storage and processing cost in order to inform their decision on which to store and which to recompute. 

In order to support the decision between caching and recomputation, the overlap across different pipelines has to be determined.
In particular, this includes the pipeline definition consisting of individual components, data sources, parameter configurations, and the general data flow.
Furthermore, it requires information about the runtime environment and resource consumption data from past pipelines.

\subsection*{Group C (\Copy{ucg:analysis}{Recommendation})}

Extensive knowledge about how to properly build pipelines can oftentimes only be acquired through years of experience.
An extensive collection of provenance information can help to unearth this knowledge.
Especially for scientists new to the field, this may save valuable time and give them a head start in their work.
In this group of UCs we explore how provenance information can be used to support the creation of pipelines.

\paragraph*{UC 5 (\Copy{uc:classification}{Recommendation of Pipeline Components})}

\noindent
Each pipeline component provides a particular functionality, e.g., removing noise from data or determining the period of a pulsar.
But a given functionality might be served by different pipeline components (including different versions thereof).
Analysing the context in which functions are used allows the identification of common functionalities provided by multiple components.
Similarly, on the pipeline level different combinations of components might serve the same purpose.
A large-scale analysis of already run pipelines may result in a taxonomy of components and pipelines.
For example, by sharing references to similar pipeline components to developers to explore alternatives for their current approach.

\underline{\textit{Scenario:}}
An astronomer is implementing a pipeline to detect objects within a very densely populated region in the Galactic Plane.
The pipeline contains a deblending process to separate nearby objects in combination with a source detection process to distinguish them. 
The pipeline returns anomalous results which the astronomer suspects to be due to the inability of the chosen processes to function in crowded fields.
The astronomer queries the provenance database to find examples of these processes applied to crowded field data.

To fulfil this requirement, at least information about components from other pipelines as well as their dataflow is needed.
Other characteristics like parameter configurations and metadata from both data sources and intermediate results might further improve the results.
Similarly, knowing the operations applied by other pipelines to the very data source can hint towards possible suggestions.
Finally, a quality metric on any result, intermediate and final, can help an implementation to rank among different possible recommendations.

\paragraph*{UC 6 (\Copy{uc:param}{Recommendation of Suitable Parameter Configurations})}

\noindent
For inexperienced users it is usually difficult to determine which parameter configurations are suitable for a given pipeline component.
The root cause of this is oftentimes an incomplete (or not up-to-date) software documentation of one or multiple pipeline components.
Collected provenance information about past pipeline runs (including their parameter configurations) can mitigate the amount of wasted computing and human resources due to repeated trial-and-error attempts to find a suitable parameter configuration.

\underline{\textit{Scenario:}} An astronomer is building a pipeline to detect objects within an image and is unsure what to set for the threshold brightness which signifies a detection. They notice that the units for this threshold are scaled to the standard deviation of the variation present within the image. The user queries past pipelines to determine what value is typically attributed to this property. 

The prime requirement to suggest possible parameter configurations is to identify components across pipelines and establish the parameter configurations used there.
Additional benefits can be gained from leveraging more context information like data sources used or metadata from both data sources and intermediate results.
Since multiple parameter configuration candidates can be extracted from previous pipeline runs, giving preference to parameter configurations that lead to high-quality results (where the quality metric is application-specific and has to be captured as additional provenance information) is preferable.
Another optimisation criteria for a parameter configuration might be the resource consumption of a particular component which requires the corresponding data for past invocations of that component.

\paragraph*{UC 7 (\Copy{uc:similarity}{Finding Similar Pipelines})}

Comparing the results of a pipeline allows them to be put into perspective.
Especially for newly created pipelines, this can provide a measure of whether the changes made improve over the current state of the art.
However, finding other pipelines, that can serve as a baseline for such an evaluation, requires an in-depth knowledge of the field at large.
When provenance records capture the inner structure of pipelines, this data can be used to define a proper similarity metric that can identify related pipelines and thus provide the grounds for further analysis.

\underline{\textit{Scenario:}}
An astronomer creates a pipeline to measure the orbital period of a binary star system. The pipeline is composed of three main steps, source detection, photometric measurement, and period determination. The results the astronomer produces appear to be inconsistent with that measured by other astronomers and they wish to determine which step in the pipeline could be the source of the discrepancy. The astronomer compares the intermediate data products of each of the three steps to their counterparts produced by other astronomers with different pipelines analysing the same object. 

Useful similarity metrics need to characterise individual pipelines in order to compare them.
In particular, this includes the components used and the data flow among them.
In addition, data sources, their metadata, and parameter configurations of used components can improve the accuracy of such a metric.
Finally, explicit data about pipeline relations given by links among pipeline versions serve as a low-effort but rather limited alternative to a generic metric.

\subsection*{Group D (\Copy{ucg:anomaly}{Anomaly Detection})}

\noindent
The final group of UCs is concerned with situations when pipelines break or behave in an unexpected manner.
Here, provenance data allows the comparison of peculiar pipeline runs with others in an attempt to investigate possible reasons and effects.

\paragraph*{UC 8 (\Copy{uc:sourceDetection}{Determining Differences Between Pipeline Executions})}

Pipelines are subject to constant evolution.
New parameter configurations are tested, components are exchanged, or the pipeline is moved to a new infrastructure.
Any of these changes might cause the pipeline to break and stop functioning.
Using provenance data of past, successful runs of other versions of the same pipeline allows the determination of differences and subsequently the identification of likely causes. 

\underline{\textit{Scenario:}}
An astronomer creates a pipeline and tests it on their machine.
They send it to a colleague to repeat the experiment on their machine but their pipeline run produces different results.
The astronomer determines all differences in runtime environments and component versions to deduce the source of the discrepancy.

The source of the anomaly is unknown and could potentially stem from any change between the involved pipelines.
So all details that impact the pipeline function are required to be contained within the provenance.
This includes all components involved, the data flow among them, data sources and parameters they consume, and information about the runtime environment.
Furthermore, to find prior pipelines to make a comparison, links to their versions, and individual runs are required.

\paragraph*{UC 9 (\Copy{uc:anomaly}{Anomaly Detection in Repeated Pipeline Runs})}

Selected pipelines are repeatedly applied on the different datasets or the latest data provided by various observatories.
As the stream of data is constantly pouring in, execution of pipelines is automated to a large extent.
With no humans directly involved, automated systems have to monitor individual executions and flag unexpected behaviour, so a human can examine it and determine possible actions.
However, what exactly constitutes \enquote{unexpected behaviour} is oftentimes hard to define.
Here, provenance can provide a baseline of common characteristics that past executions of a specific pipeline or individual components exhibited.
Instead of profiling pipeline components in selected configurations in advance, a comprehensive provenance database offers a detailed picture of regular operations.
This also allows adaptation to continuous shifts in performance automatically as caused by, e.g., adding new hardware or changes in telescope characteristics.

\underline{\textit{Scenario:}}
An astronomer builds a pipeline to measure the impact of the atmosphere on astronomical images. One of the key metrics in this determination is the point spread function (PSF) - a measure of how large a point source appears in an image. The PSF is stored as a cutout image which represents the average point source shape within a single image. The astronomer wishes to be notified if the PSF for any single region of sky changed significantly over its observations. The astronomer programs the system such that it will notify them if the physical size of the PSF image exhibits significant change. 

The minimum requirements are the resource consumption and data product characteristics of the individual pipeline in all its versions.
The context of a particular pipeline run is given by its runtime environment, data source metadata, and parameter configurations.
Furthermore, it requires information about other versions of the considered pipeline and their corresponding runs.
In an extension, the black-box-model of the pipeline can be replaced and individual components be considered.
Information previously only required on the pipeline level, now needs to be available for each component and their intermediate results.

\subsection*{Summary}
\label{sect:reqs}

\newcommand{\man}[1][]{\(\CIRCLE_{#1}\)}     
\newcommand{\opt}[1][]{\(\LEFTcircle_{#1}\)} 
\newcommand{\non}{\Circle}                   

\begin{table*}[t]
    \centering
    \begin{tabular}{cc|cc|cc|ccc|cc}

                            & & \multicolumn{2}{c}{Group A}     & \multicolumn{2}{c}{Group B} & \multicolumn{3}{c}{Group C} & \multicolumn{2}{c}{Group D} \\
                            &                                   & \rot{UC 1} & \rot{UC 2} & \rot{UC 3} & \rot{UC 4} & \rot{UC 5} & \rot{UC 6} & \rot{UC 7} & \rot{UC 8} & \rot{UC 9} \\
\hline
\multirow{6}{*}{Identifiers}
                            & Pipelines
  & \non & \non & \opt & \non & \non & \non & \opt & \man & \man \\
                            & Pipeline Runs
  & \man & \non & \opt & \non & \non & \non & \non & \man & \man \\
                            & Components
  & \man & \man & \man & \man & \man & \man & \man & \man & \opt \\
                            & Data Sources
  & \man & \man & \non & \man & \opt & \opt & \opt & \man & \non \\
                            & Data Products
  & \non & \man & \non & \non & \non & \non & \non & \non & \non \\
                            & Intermediate Results
  & \non & \non & \non & \non & \non & \non & \non & \non & \non \\
\hline
\multirow{7}{*}{Attributes}
                            & Parameters
  & \man & \man & \opt & \man & \opt & \man & \opt & \man & \man \\
                            & Runtime Environment
  & \opt & \non & \opt & \man & \non & \non & \non & \man & \man \\
                            & Resource Consumption
  & \non & \non & \man\textsubscript{a} & \man & \non & \opt & \non & \non & \man \\
                            & Data Source Metadata
  & \non & \opt & \opt & \non & \opt & \opt & \opt & \non & \non \\
                            & Data Product Metadata
  & \non & \non & \non & \non & \non & \non & \non & \non & \man \\
                            & Interm. Result Metadata
  & \non & \non & \non & \man & \opt & \opt & \non & \non & \opt \\
                            & Quality Metrics
  & \non & \non & \man\textsubscript{a} & \non & \opt & \opt & \non & \non & \non \\
\hline
\multirow{2}{*}{Connections} 
                            & Data Flow
  & \man & \man & \opt & \man & \man & \opt & \man & \man & \non \\
                            & Pipeline Version
  & \non & \non & \opt & \non & \non & \non & \opt & \man & \man \\
\hline
\multirow{2}{*}{Prov. Records} & same Pipeline
  & \man & \non & \non & \non & \non & \non & \non & \non & \non \\
                            & other Pipelines
  & \non & \man & \man & \man & \man & \man & \man & \man & \man \\
\hline
\hline
\multirow{3}{*}{Other UCA} & Miles et. al. \cite{miles2007requirements}
  & \man & \man & \man & \non & \non & \man & \non & \man & \man \\
                            & Bowers et al. \cite{bowers2006model}
  & \man & \non & \non & \non & \non & \non & \non & \non & \man \\
                            & Ram et al. \cite{ram2009new}
  & \man & \man & \non & \non & \non & \non & \non & \non & \non \\
\end{tabular}

    \caption{
        Summary of requirements per use case. 
        \man{} \ldots{} mandatory; \opt{} \ldots{} optional; \non{} \ldots{} not required.
        \\
        For entries with the same subscript, at least one requirement has to be fulfilled.
        The "Other UCA" rows denote whether each use case was included in other analyses (further discussion in Section \ref{sec:RW}). 
        \man{} \ldots{} included; \non{} \ldots{} not included.
    }
    \label{tab:req_summary}
\end{table*}

\autoref{tab:req_summary} summarises the requirements posed by the different use cases.
For better readability, we grouped the requirements into the following five categories:

\setlist[description]{font=\normalfont\itshape\space}
\begin{description}

 \item[Identifiers] provide direct access to specific parts of a pipeline.
    Otherwise, the graph structure has to be traversed to retrieve a particular node.
    In particular, identifiers allow to relate components across the borders of a single pipeline.
    
  \item[Attributes] capture additional information about the pipeline in general or parts thereof.
  
  \item[Connections] allow to model the relationship among parts of a provenance record.
  
  \item[Provenance Records] defines which records need to be accessed in order to fulfil a given use case.
    In particular, this illustrates whether an individual provenance record is sufficient or a larger provenance database is required.

\end{description}

We further aggregate the requirements of individual use cases into broader classes.
We distinguish between parameters and (intermediate) datasets.
While the former denote values used to configure a particular component like a threshold for a filter, the latter is generally comprised of larger amounts of data that are stored in files.
The runtime environment can be characterised along different dimensions.
In the context of this analysis, we adopt a definition that includes all system characteristics that can be determined automatically like the version of the operating system, available CPU cores, or size of accessible memory.
Similarly, resource consumption encompasses the share of available resources that are actually used by a given pipeline or its component.

In the same vein, metadata in this context is mostly comprised of information that can automatically be derived from a dataset.
Of particular interest are aspects like file size, contained entries, and other, rather technical characteristics.
Aspects like title, description, or creator are mostly used for human consumption and are thus largely out of scope here.
We also introduce the quality of a result as an additional criterion.
This can replace inferred quality measures (e.g., frequency of use) with an explicit metric able to capture domain-specific aspects that may otherwise go unnoticed.



The most common requirements across all use cases are identifiers for components, call parameters, and the data flow between components.
Identifiers for data sources however, are required slightly less often. 


On the other end of the spectrum, data product identifiers and corresponding metadata are only required once each.
While identifiers for data products are necessary to report those that are affect by some kind of error (UC2), metadata for data products is needed to distinguish common from unusual pipeline behaviour (UC9). Prior to the study, we assumed that identifiers for intermediate results might be of use at least in some use cases.
However, during discussion it became evident that it is entirely sufficient to refer to them as part of the data flow and not on their own.
Metadata on intermediate products is only necessary in one use case (\Paste{uc:storage} - UC4).
This raises the question whether tracking information on that level of details is actually worth the cost of storing it to begin with.

The final observation concerns the necessity to have access to a broad range of provenance records in most use cases.
With the exception of re-executing on particular pipeline run (UC1), all other use cases require access to provenance records from a variety of pipelines and their runs.

\section{Opportunities and Challenges}\label{sec:OandC}

The use case analysis in \autoref{sec:usecases} and the requirements derived from it can be combined to outline the main challenges and opportunities for recording the provenance from astronomical pipelines.
This discussion can also be extended by considering the expected data scales created by future astronomical survey telescopes and the likely composition of the pipelines built to analyse these data.
The challenges and opportunities derived from each of these sources are collated within this section.   

\paragraph*{The Scale of Astronomical Datasets} The expected exabyte scale datasets within astronomy means that it shares the challenge of scalable and distributed provenance recording, along with many other data intensive applications.
To keep up with the rapid growth of astronomical datasets, there has also been a rapid evolution of astronomical processing techniques.
This necessitates that the provenance that describes it must also be able to adapt and be compatible with provenance from past pipeline versions.
The scale of astronomical data also means that these applications have the potential to provide a rich source of provenance data. 

\paragraph*{Interoperability of Provenance Records}
Almost all presented use cases rely on an extensive collection of provenance records from different pipelines.
This dependence on data from different origins highlights the necessity to make this data interoperable.
Systems need to be able to recognise similarities across pipelines and pipeline runs regardless of their particular origin.
This goes beyond the coarse classification of W3C's PROV standard \cite{w3c_prov_primer} and calls for domain specific extensions.

\paragraph*{Provenance for Knowledge Sharing} Groups B and C detail the use of provenance to support researchers that are not (yet) experts in all aspects of astronomical pipelines.
Provenance can capture the wisdom of said experts that usually takes years over years of experience.
Sharing that knowledge will benefit researchers new to the field and thus the community at large.
However, first and foremost this requires an open culture of sharing in the domain.
Without such a free exchange of information, the use cases proposed here will not be able to unfold their full potential.
These use cases also rely upon an intent of component function which must be generalisable in order to identify similar functions across pipelines yet specific enough to only match compatible functions.
The determination of this granularity and the corresponding implementation both present challenges.  


\paragraph*{The Addition of Quality Metrics} Quality and provenance have been explored in past works in ways such as determining the quality of the pipeline from the provenance or determining the quality of the provenance itself.
However, adding quality metrics that describe the pipeline to the provenance is previously unexplored and can offer new opportunities such as those described within UCs 3, 5, and 6.
The calculation of the quality itself will likely be different for each pipeline and is required to be generated by the user.
The challenge from the provenance perspective will be to facilitate the addition of this quality post-processing and the determination of whether quality metrics would be most useful if describing the pipeline as a whole or of specified portions.  

\section{Related Work}\label{sec:RW}


The analysis of provenance use cases within the scientific domain is discussed by Miles et al. \cite{miles2007requirements} in which they outline a number of generalised use cases within the domains of chemistry, biology, physics, and computer science.
They proceed to discuss the implications that solving these use cases has on the architecture of potential provenance generation systems and describe a provenance model for e-science experiments. 

Bowers et al. \cite{bowers2006model} describe a formal approach to generalisable use cases for scientific pipelines within the domain of life sciences.
The main focus of these use cases was identifying the relationships between and characteristics of the data processing and associated data products. 

Ram et al. \cite{ram2009new} investigated the domains of biology, business and physical sciences and outline use cases within these fields which are aimed to establish reliability and reproducibility. 




\section{Conclusions and Future Work}\label{sec:conc}

In this paper, we present a number of generalisable use cases for provenance within the context of astronomical data processing.
From this use case analysis, we derived the requirements that solving these use cases would impose on the provenance system.
Furthermore, we discussed the opportunities and challenges associated with building such a system for astronomical pipelines.
This work represents the first steps of the VAMPIRA project which aims to facilitate the automatic generation of provenance for astronomical pipelines.

The next step for the VAMPIRA project is to use this use case analysis  as the basis to construct a model which accommodates all the needs of the astronomer for provenance generation.
Subsequently, a tool will be created to automatically generate the provenance from Python scripts in an accessible and scaleable manner.
The process for the generation of both the tool and provenance model will include an evaluation of the requirements stated in this paper against both existing models and tools.
This tool will be used to record the provenance from the two pipelines which motivated the use case analysis - the EPTA pipeline and Kepler pipeline.
Furthermore, the suitability of the tool will be evaluated by solving application specific scenarios for each described use case within each pipeline. 
The suitability of the tool will be furhter applied to another pipeline designed to analyse data from the MeerKAT radio telescope. As this pipeline was not a motivator for the use case analysis, this should provide a test of the generalisability of the constructed provenance model.

The discussion within this paper is limited to use cases within astronomy.
In the future, the work for VAMPIRA will expand to include data intensive applications within other scientific domains such as remote sensing programmes and upcoming satellite programmes to explore/monitor the Earth ecosystem.

In conclusion, by outlining use cases for astronomical provenance we have laid the foundation for the generation of provenance within astronomy.
Our goals are to build from this foundation and enable the automated provenance generation for pipelines within astronomy and, in future, other data intensive scientific disciplines. 

\section{Acknowledgements}

Supported by the German Federal Ministry for Economic Affairs and Energy on the basis of a decision of the German Bundestag under the project number 50OO1905.

{\footnotesize \bibliographystyle{acm}
\bibliography{sample}}

\begin{thebibliography}{10}

\bibitem{altintas2006provenance}
{\sc Altintas, I., Barney, O., and Jaeger-Frank, E.}
\newblock Provenance collection support in the kepler scientific workflow
  system.
\newblock In {\em International Provenance and Annotation Workshop\/} (2006),
  Springer, pp.~118--132.

\bibitem{w3c_prov_primer}
{\sc Belhajjame, K., Deus, H., Garijo, D., Klyne, G., Missier, P.,
  Soiland-Reyes, S., and Zednik, S.}
\newblock {PROV Model Primer}.
\newblock \url{https://www.w3.org/TR/prov-primer/}.

\bibitem{bowers2006model}
{\sc Bowers, S., McPhillips, T., Lud{\"a}scher, B., Cohen, S., and Davidson,
  S.~B.}
\newblock A model for user-oriented data provenance in pipelined scientific
  workflows.
\newblock In {\em International Provenance and Annotation Workshop\/} (2006),
  Springer, pp.~133--147.

\bibitem{foster2002chimera}
{\sc Foster, I., Vockler, J., Wilde, M., and Zhao, Y.}
\newblock Chimera: A virtual data system for representing, querying, and
  automating data derivation.
\newblock In {\em Proceedings 14th International Conference on Scientific and
  Statistical Database Management\/} (2002), IEEE, pp.~37--46.

\bibitem{jaeger2008common}
{\sc Jaeger, S.}
\newblock The common astronomy software application (casa).
\newblock In {\em Astronomical Data Analysis Software and Systems XVII\/}
  (2008), vol.~394, p.~623.

\bibitem{juric2015lsst}
{\sc Juri{\'c}, M., Kantor, J., Lim, K., Lupton, R.~H., Dubois-Felsmann, G.,
  Jenness, T., Axelrod, T.~S., Aleksi{\'c}, J., Allsman, R.~A., AlSayyad, Y.,
  et~al.}
\newblock The lsst data management system.
\newblock {\em arXiv preprint arXiv:1512.07914\/} (2015).

\bibitem{lazarus2020coastguard}
{\sc Lazarus, P., Karuppusamy, R., Graikou, E., Caballero, R., Champion, D.,
  Lee, K., Verbiest, J., and Kramer, M.}
\newblock Coastguard: Automated timing data reduction pipeline.
\newblock {\em Astrophysics Source Code Library\/} (2020), ascl--2003.

\bibitem{mcphillips2015yesworkflow}
{\sc McPhillips, T., Song, T., Kolisnik, T., Aulenbach, S., Belhajjame, K.,
  Bocinsky, K., Cao, Y., Chirigati, F., Dey, S., Freire, J., et~al.}
\newblock {YesWorkflow}: A user-oriented, language-independent tool for
  recovering workflow information from scripts.
\newblock {\em International Journal of Digital Curation 10}, 1 (may 2015),
  298--313.

\bibitem{miles2007requirements}
{\sc Miles, S., Groth, P., Branco, M., and Moreau, L.}
\newblock The requirements of using provenance in e-science experiments.
\newblock {\em Journal of Grid Computing 5}, 1 (2007), 1--25.

\bibitem{murta2014noworkflow}
{\sc Murta, L., Braganholo, V., Chirigati, F., Koop, D., and Freire, J.}
\newblock noworkflow: capturing and analyzing provenance of scripts.
\newblock In {\em International Provenance and Annotation Workshop\/} (2014),
  Springer, pp.~71--83.

\bibitem{oinn2004taverna}
{\sc Oinn, T., Addis, M., Ferris, J., Marvin, D., Senger, M., Greenwood, M.,
  Carver, T., Glover, K., Pocock, M.~R., Wipat, A., et~al.}
\newblock Taverna: a tool for the composition and enactment of bioinformatics
  workflows.
\newblock {\em Bioinformatics 20}, 17 (jun 2004), 3045--3054.

\bibitem{ram2009new}
{\sc Ram, S., and Liu, J.}
\newblock A new perspective on semantics of data provenance.
\newblock {\em SWPM 526\/} (2009).

\end{thebibliography}


\end{document}